\documentclass[english,letterpaper,prl,aps,showpacs,superscriptaddress,floatfix,twocolumn]{revtex4}
\usepackage[T1]{fontenc}
\usepackage[latin9]{inputenc}
\usepackage{verbatim}
\usepackage{amstext}
\usepackage{amssymb}
\usepackage{graphicx}

\makeatletter

\@ifundefined{textcolor}{}
{%
 \definecolor{BLACK}{gray}{0}
 \definecolor{WHITE}{gray}{1}
 \definecolor{RED}{rgb}{1,0,0}
 \definecolor{GREEN}{rgb}{0,1,0}
 \definecolor{BLUE}{rgb}{0,0,1}
 \definecolor{CYAN}{cmyk}{1,0,0,0}
 \definecolor{MAGENTA}{cmyk}{0,1,0,0}
 \definecolor{YELLOW}{cmyk}{0,0,1,0}
}

\makeatother

\usepackage{babel}
\begin{document}

\title{Quantum Optomechanical Heat Engine}
\author{Keye Zhang}
\affiliation{Quantum Institute for Light and Atoms, Department of Physics, East
China Normal University, Shanghai, 200241, People's Republic of China }
\affiliation{B2 Institute, Department of Physics and College of Optical Sciences,
University of Arizona, Tucson, Arizona 85721, USA}
\author{Francesco Bariani}
\affiliation{B2 Institute, Department of Physics and College of Optical Sciences,
University of Arizona, Tucson, Arizona 85721, USA}
\author{Pierre Meystre}
\affiliation{B2 Institute, Department of Physics and College of Optical Sciences,
University of Arizona, Tucson, Arizona 85721, USA}
\begin{abstract}
We investigate theoretically a quantum optomechanical realization of a heat engine. In a generic optomechanical arrangement the optomechanical coupling between the cavity field and the oscillating end mirror results in polariton normal mode excitations whose character depends on the pump detuning and the coupling strength. By varying that detuning it is possible to transform their character from phononlike to photonlike, so that they are predominantly coupled to the thermal reservoir of phonons or photons, respectively. We exploit the fact that the effective temperatures of these two reservoirs are different to produce an Otto cycle along one of the polariton branches. We discuss the basic properties of the system in two different regimes: in the optical domain it is possible to extract work from the thermal energy of a mechanical resonator at finite temperature, while in the microwave range one can in principle exploit the cycle to extract work from the blackbody radiation background coupled to an ultracold atomic ensemble.
\end{abstract}
\maketitle

Heat engines operating in the quantum regime have attracted much recent attention due to their potential to investigate the quantum limit of classical thermodynamical concepts such as the Carnot efficiency limit---and perhaps  overcome that limit, to better understand thermalization in the deep quantum regime, and, on a more applied side, in the quest for the realization of nanoengines of increasingly small size~\cite{Scully02, Linden10,RoBFnagel14,Horodecki13,Correa14}. Microscopic scale heat engines have been realized in micro-electro-mechanical systems~\cite{ Whalen03, Steeneken11}, but reaching the quantum regime remains a challenge due to thermal noise in the mechanical elements. Theoretical proposals for quantum heat engines have been advanced involving single ions~\cite{Abah12}, ultracold bosonic atoms~\cite{Fialko12}, and quantum dots~\cite{dotheatengine}, but so far their experimental realization has remained elusive.

This Letter proposes and analyzes theoretically a quantum heat engine based on a cavity optomechanical setup. This system presents several attractive features: first, it is a truly mechanical system; second, it has the potential to operate deep in the quantum regime using existing, state-of-the-art equipment; third, it is conceptually extremely simple; and fourth,  it offers, in principle at least, the potential  to extract work from the 2.7~K blackbody radiation background. Finally, when combined with progress in quantum optics toward the realization of squeezed reservoirs \cite{Kimble}, it may provide a route to testing the Carnot efficiency limit in the quantum regime. 

The key element of a heat engine is a medium that may be used to extract work and that exchanges heat with thermal reservoirs at two different temperatures.  Cavity optomechanics provides a conceptually simple way to realize that goal: The radiation pressure force permits the exchange of energy  between cavity photons and mechanical phonons, and crucially the cavity and mechanical damping couple the system to both a cold and a hot reservoir. Cavity optomechanics has witnessed spectacular developments in the last decade (see, e.g.~Refs~\cite{RMP,Meystre,DSK} for recent reviews), and can operate deep in the quantum regime~\cite{Oconnell,Teufel,Painter}.   Also, quantum entanglement and squeezed states of photons and phonons have been demonstrated in these systems~\cite{Palomaki13,Naeini13}.

We consider a standard optomechanical setup with a cavity mode at frequency $\omega_{c}$ coupled to a mechanical resonator at frequency $\omega_{m}$ for example, the harmonically bound end mirror of a Fabry-P{\'e}rot resonator, with single-photon coupling strength $g$. The resonator is driven by an optical pump field with strength $\alpha_{\rm in}$ and frequency $\omega_p$.  In addition, the intracavity field and mechanical oscillator suffer damping of rates $\kappa$ and $\gamma$.
We assume that the intracavity field is strong enough that it can be described as the sum of a large mean field $\alpha$ and small quantum fluctuations. In a frame rotating at $\omega_p$ the Hamiltonian of the entire system can then be linearized as 
\begin{equation}
H=-\hbar\Delta\hat{a}^{\dagger}\hat{a}+\hbar\omega_{m}\hat{b}^{\dagger}\hat{b}+\hbar G(\hat{b}+\hat{b}^{\dagger})(\hat{a}+\hat{a}^{\dagger}),
\end{equation}
where the bosonic annihilation operators $\hat{a}$ and $\hat{b}$ account for the fluctuations of the photon and phonon mode annihilation operators around their mean amplitudes $\alpha$ and $\beta$,  $G=\alpha g$ and the detuning $\Delta=\omega_{p}-\omega_{c}-2\beta g$ includes the mean radiation pressure induced change in resonator length. In steady state  $\alpha=\alpha_{\rm in}/\Delta$ and $\beta=-g\alpha^2/\omega_m$ for small damping~\cite{steady}. (We take $\alpha$ and $\beta$ to be real without loss of generality in the following.) The quadratic Hamiltonian $H$ describes two linearly coupled harmonic oscillators. In the red detuned regime $\Delta<0$, it  leads in general to stable dynamics that can result in sideband cooling \cite{cooling,cooling limit}. 

\begin{figure}
\includegraphics[scale=0.5]{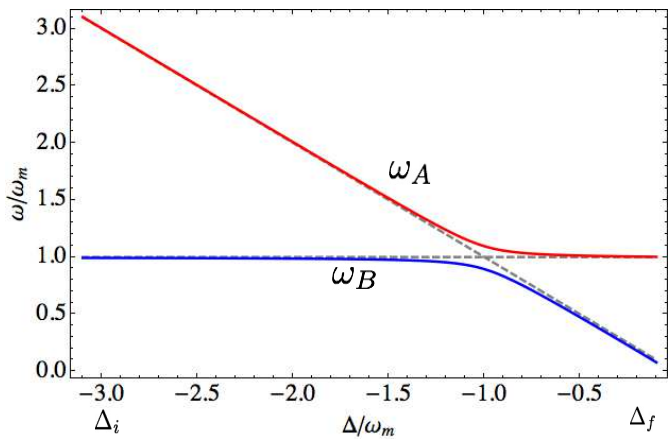}
\caption{(Color online). Eigenfrequencies of the normal modes $A$ and $B$, in units of $\omega_m$, as functions of the normalized cavity detuning $\Delta/\omega_{m}$ for the dimensionless optomechanical coupling strength $G/\omega_{m}=0.1$. Dashed lines: noninteracting energies of the phonon and photon modes. $\Delta_i$ and $\Delta_f$ are the initial and final detunings for a generic Otto cycle. \label{fig:Two-branches}}
\end{figure}

To discuss the energy conversion between photons and phonons it is convenient to introduce a normal mode representation of the system. After removing a constant term, we can express $H$ in the diagonal form \cite{Brokje13, Lemonda13} 
\begin{equation}
H=\hbar\omega_{A}\hat{A}^{\dagger}\hat{A}+\hbar\omega_{B}\hat{B}^{\dagger}\hat{B},
\end{equation}
where the new operators $\hat{A}$ and $\hat{B}$ are the boson annihilation operators for the normal-mode excitations of the system (polaritons), with frequencies
\begin{equation}
\omega_{A,B}=\sqrt{\frac{\Delta^{2}+\omega_{m}^{2}\pm\sqrt{(\Delta^{2}-\omega_{m}^{2})^{2}-16G^{2}\Delta\omega_{m}}}{2}}.\label{eq:spectrum}
\end{equation}
In general, these excitations are superpositions of the cavity field and the mechanics. As shown in Fig.~\ref{fig:Two-branches}, for the sideband resonant case $\Delta=-\omega_{m}$, the degeneracy in the uncoupled energy spectrum is lifted by the optomechanical interaction and normal mode splitting occurs with a separation of the order of $2G$, as been experimentally observed in Ref.~\cite{Aspelmeyer09}.

For $\Delta\ll-\omega_{m}$, the low-energy polariton branch, characterized by the bosonic annihilation operator $\hat B$ and the frequency $\omega_B(\Delta)$, describes phononlike excitations, with $\omega_{B}$ approaching $\omega_{m}$. In contrast, on the other side of the avoided crossing, $-\omega_{m}\ll\Delta<0$, and in the weak coupling regime $G/\omega_m \ll 1$, the operator $\hat{B}$ annihilates photon-like excitations of frequency $\omega_B \sim -\Delta$. The opposite holds for the polariton branch of frequency $\omega_A(\Delta)$, which is photonlike for frequencies far red-detuned from $\Delta = -\omega_m$ and phonon-like on the other side.

In addition to the coherent dynamics, these excitations also undergo damping and decoherence,  resulting in the thermalization of the system. The polariton decay rates $\Gamma_A$ and $\Gamma_B$ are combinations  of the cavity decay rate $\kappa$ and mechanical damping rate $\gamma$ \cite{Brokje13}, the temperatures of the associated thermal reservoirs $T_A$ and $T_B$ depending on the original bath temperatures, $T_{a}$ for the photons and $T_{b}$ for the phonons. At optical frequencies it is an excellent approximation under normal laboratory conditions to take $T_a \approx 0 K$---but as we discuss later on, this is not the case in the microwave regime. We then have $T_a \ll T_{b}$. Both the properties of the normal-mode excitations, and thus their photonlike or phononlike nature, and their reservoir temperatures  are controlled by the detuning $\Delta$. The proposed heat engine relies on this simple observation: it operates by varying $\Delta$ to cycle the nature of the polariton between phononlike and photonlike and exploits the difference in the associated effective reservoir temperatures to extract work from the system. 

We proceed by first considering a quantum heat engine that operates along a single polariton branch. We focus specifically on the lower energy normal-mode  $B$ and consider an Otto cycle \cite{td_book} consisting of the following four consecutive steps.

(1) \emph{Isentropic expansion}.---This step is achieved by varying the detuning from its initial value $\Delta_{i}\ll-\omega_{m}$, where the polariton is to an excellent approximation phononlike, to the final value $-\omega_{m}\ll\Delta_{f}<0$ over a time interval $\tau_{1}$. In this step  $\omega_B$ changes from the high value $\omega_{i}=\omega_B(\Delta_i)$ to a lower frequency $\omega_{f}=\omega_B(\Delta_f)$. The change in $\Delta$ should occur in such a way that the mean intracavity optical field amplitude $\alpha$ remains constant.  In addition the speed of the process must be such that two potentially conflicting requirements are simultaneously satisfied. First, it must be fast enough to be very nearly isentropic: such transformations are carried out by thermally insulating the system from its reservoirs, so that the thermal mean particle number $\bar{N}_{i}=\langle \hat{B}^{\dagger}\hat{B}\rangle _{\omega_{i},T_{i}}$ at the initial temperature $T_{i}$ and frequency $\omega_{i}$ remains unchanged. Since the coupling to the thermal reservoirs cannot be switched off in our optomechanical system, we must therefore have that $\tau_{1}$ is  short compared to the phonon thermalization time and the cavity decay time. This can, however, conflict with a second requirement that the transformation be slow enough to be adiabatic \cite{note}, in the sense that the system does not undergo transitions between the two polariton branches. This requires that $1/\tau_{1}$ be much smaller than the smallest frequency separation between the excitation bands $A$ and $B$, which occurs at $\Delta = -\omega_m$ and is of order $2G$.

(2) \emph{Cold isochoriclike transition}.---At this point the photonlike polariton $B$ is predominantly coupled to the photon reservoir at temperature $T_{f}\approx 0$ K, and is allowed to thermalize over a time $\tau_{2}$, the detuning remaining fixed at the value $\Delta_f$. During that step, whose duration must be  $1/\tau_{2} < \kappa$ to ensure full thermalization, the thermal occupation adjusts to a lower thermal mean particle number, $\bar{N}_{i} \rightarrow \bar{N}_{f}$. 

(3) \emph{Isentropic compression}.---The detuning $\Delta$ is then returned to its initial value, during which step the polariton frequency returns to the phononlike higher value $\omega_{i}$ with $\bar{N}_{f}$ remaining constant provided that $\tau_{3}$  satisfies the same conditions as $\tau_{1}$. 

(4) \emph{Hot isochoric-like transition}.---The polariton,  its frequency now fixed at $\omega_{i}$, remains in contact with the phonon reservoir for a time $1/\tau_{4} < \gamma$, and its thermal population returns to the initial value $\bar{N}_{i}$. 

\begin{figure}
\includegraphics[width=9.cm,height=6.5cm]{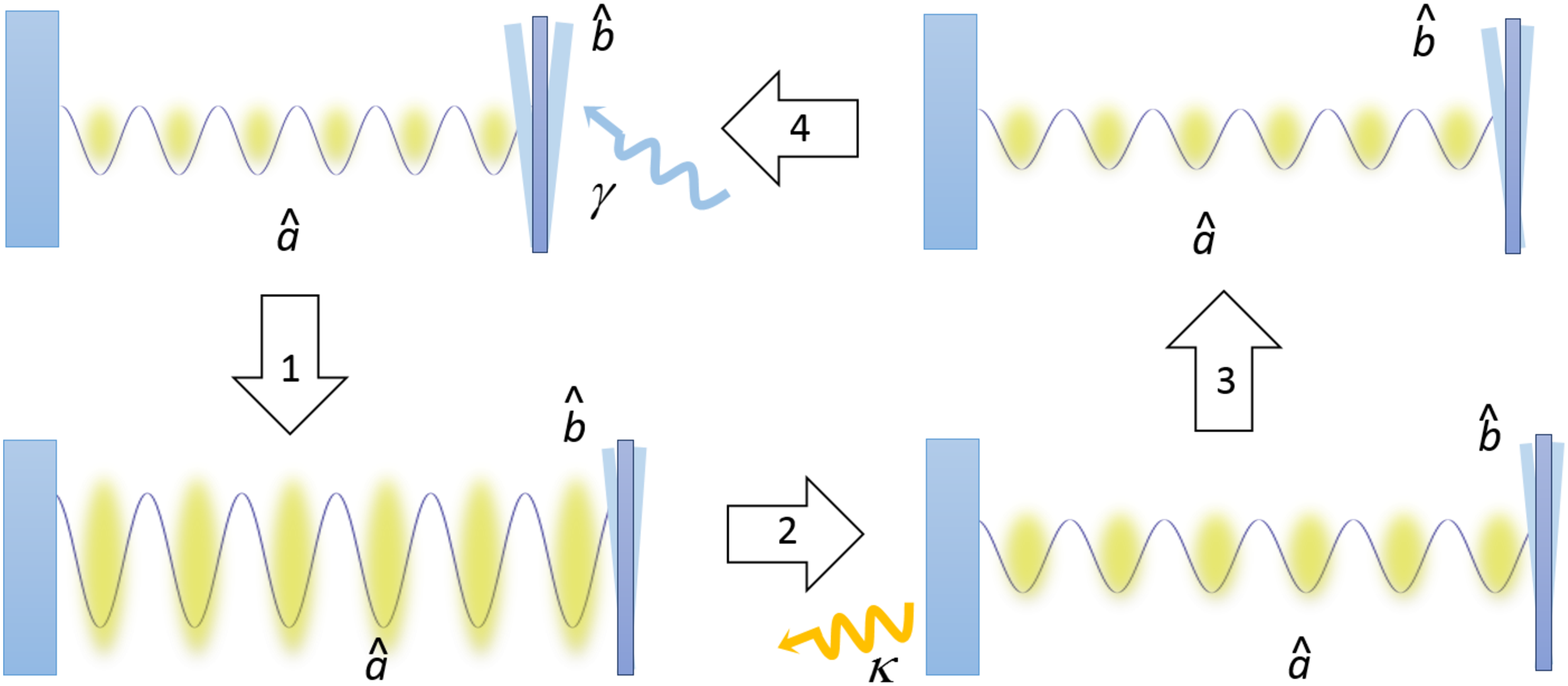}
\caption{(Color online). Intuitive physical picture of the Otto cycle for the optomechanical heat engine, see text for details.\label{fig:cycle}} \end{figure}

One can gain a simple physical understanding of the engine cycle by considering the effects of varying the detuning $\Delta$, see Fig.~\ref{fig:cycle}. In practice, this can be achieved by changing the frequency of the driving field, but importantly, we emphasize that all detuning changes must be performed while simultaneously changing the pumping rate $\alpha_{\rm in}$ so that the mean intracavity amplitude $\alpha$ remains constant. 

During stroke $(1)$  $\Delta$ is varied, so that $\omega_p$  becomes closer to resonance with the cavity mode frequency $\omega_c$. As this happens the phononlike thermal excitations, which are initially large due to the contact with a thermal bath that is essentially at the temperature of the mechanics, are transformed into photonlike excitations. This occurs at a rate characterized by the coupling strength $G$. In this step the vibration amplitude of the mechanical resonator decreases. The excess energy is infused into the intracavity field, and as a result the resonator length increases by a small amount due to the increased radiation pressure. It is at this point that the mechanical work on the oscillator is produced by the optomechanical heat engine. During the thermalization step of stroke $(2)$  the population of the photonlike excitations decays to zero at rate $\kappa$ (for a photon reservoir at zerotemperature). If the resonator length were to instantly return to its initial position following this decay, the total mechanical work would then be zero. But if $\kappa\gg\gamma$, as is often the case in cavity optomechanics, then changes in cavity length as well as the population of the phonon-like excitations can be neglected during the time $\tau_2$. In stroke $(3)$ the remaining polariton excitations (if any) are turned back into phononlike quanta by adjusting $\Delta$. The phonon branch is finally repopulated via thermal contact with the hot mechanical reservoir in stroke $(4)$ and the cavity length also returns to its initial value.

We now analyze the performance of this heat engine, following the approach of Ref.~\cite{Abah12} to determine the total work and thermal efficiency of the Otto cycle. The average values of the energy of the system at the four stages 
are given by $E_{1}=\hbar\omega_{i}\bar{N}_{i}$, $E_{2}=\hbar\omega_{f}\bar{N}_{i}$, $E_{3}=\hbar\omega_{f}\bar{N}_{f}$, and $E_{4}=\hbar\omega_{i}\bar{N}_{f}$, and the total work per cycle is $W=E_{1}-E_{2}+E_{3}-E_{4}$.  The thermal efficiency is $\eta=W/Q$, defined as the ratio of the total work per cycle and the heat received from the hot reservoir, $Q=E_{1}-E_{4}$ which corresponds to stroke $(4)$. The
total work and heat received per cycle are
\begin{eqnarray}
W & = & \hbar(\omega_{i}-\omega_{f})(\bar{N}_{i}-\bar{N}_{f}),\\
Q & = & \hbar\omega_{i}(\bar{N}_{i}-\bar{N}_{f}),
\end{eqnarray}
where the conditions $\omega_{i}>\omega_{f}$ and $\bar{N}_{i}>\bar{N}_{f}$ ensure that $W$ and $Q$ are positive, and the thermal efficiency is 
\begin{equation}
\eta=1-\frac{\omega_{f}}{\omega_{i}}.\label{eq:eta}
\end{equation}

The total work depends on the mean polariton numbers $\bar{N}_i$ and $\bar {N}_f$, which are combinations of the thermal phonon number $\bar{n}_{b}$ and the thermal photon number $\bar{n}_{a}$. The coefficients of these combinations are given by the Bogoliubov diagonalization. Their analytical expressions are cumbersome and not very transparent, and we proceed instead with a numerical study of the main feature of the engine cycle. We choose $\Delta_{i}=-3\omega_{m}$, so that the polariton population is predominantly on the lower polariton branch $B$, and evaluate numerically  $\eta$ and $W$ as a function of the normalized coupling strength $G/\omega_{m}$ and final detuning $\Delta_f/\omega_{m}$. 

\begin{figure}
\includegraphics[width=9cm]{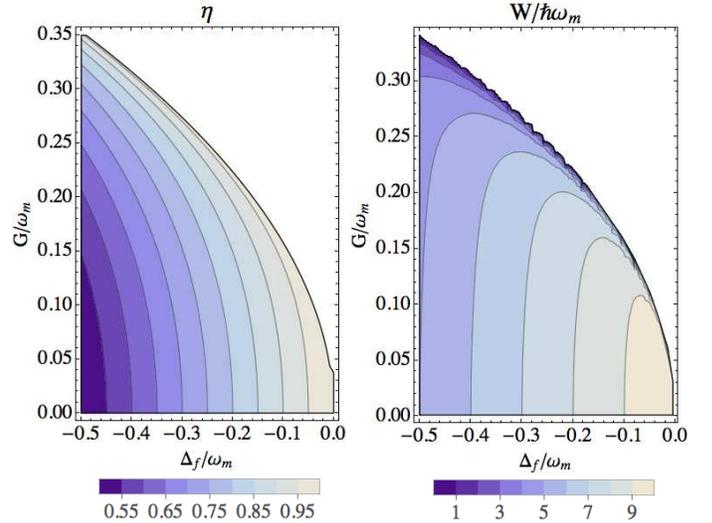}
\caption{(Color online). Thermal efficiency and total work of the Otto cycle, in units of $\hbar \omega_m$,  for $T_{a}=0$ and $T_{b}=0.1$ K, corresponding to $\bar{n}_a=0$, $\bar{n}_b=10$, and $\omega_m=200$ MHz.
In the white region the linearized system is unstable. \label{fig:The-thermal-efficient}}
\end{figure}

The results are summarized in Fig.~\ref{fig:The-thermal-efficient}, which illustrates the trade-off between  maximum work and maximum efficiency, as already discussed in previous works~\cite{Leff87}. The maximum efficiency is reached for $G/\omega_{m}=\sqrt{-\Delta_{f}/\omega_{m}}/2$, which follows from the condition $\omega_{f}=0$---note that this is also the stability threshold for the linearized form of optomechanical coupling that we consider---and the maximum amount of work is extracted for small values of $G/\omega_{m}$ and $-\Delta_f/\omega_m$.  (For large values of $G/\omega_{m}$ and $-\Delta_f/\omega_m$, the polariton branch $B$ is no longer strongly photonlike. In this case we find that while the efficiency may still be high, the output work $W$ is reduced. Note, however, that here the simple heuristic argument that we invoked to separate the effects of the two reservoirs ceases to be appropriate. )

For $(G/\omega_{m}, -\Delta_f/\omega_m) \ll1$ we can derive perturbative analytical forms for $W$ and $\eta$. In  that limit the upper and lower frequencies of the cycle are $\omega_{i}=\omega_{m}$ and $\omega_{f}=-\Delta_f-2G^{2}/\omega_{m}$, and the thermal mean polariton numbers are $\bar{N}_{i}=\bar{n}_{b}$ and $\bar{N}_{f}=(1+4\Delta_{f}G^{2}/\omega_{m}^{3})\bar{n}_{a}+2G^{2}\bar{n}_{b}/\omega_{m}^{2}$, with $\bar{n}_{a}=0$ in the case of optical frequencies and $\bar{n}_{b}=1/(e^{\hbar\omega_{m}/k_{B}T_{b}}-1)$, where $k_B$ is the Boltzmann constant.  The total work is then
\begin{equation}
\frac{W}{\hbar \omega_m} =  \left(\frac{\Delta_f}{\omega_m}+\frac{2G^{2}}{\omega_{m}^{2}}+1\right) \left[\left(1-\frac{2G^{2}}{\omega_m^2}\right)\bar{n}_{b}-\frac{G^{2}}{\omega_m^2}\right].
\end{equation}
In the high temperature limit of the phonon bath, $\hbar\omega_{m}/(k_{B}T_{b})\ll1$, $W$ is maximum for $G^{2}/\omega_m^2=-\Delta_f/(4\omega_m)-\hbar\omega_{m}/(8k_{B}T_{b})$. If we substitute this into Eq.~(\ref{eq:eta}), we obtain the efficiency at maximum work
\begin{equation}
\eta_W=1-\left (\frac{-\Delta_{f}}{\omega_{m}}+\frac{\hbar\omega_{m}}{4k_{B}T_{b}}\right ).
\end{equation}
Remembering that $\Delta_f < 0$, this shows that the efficiency is limited by
\begin{equation}
\eta_W<1-\sqrt{\frac{-\hbar \Delta_f}{2k_{B}T_{b}}},
\end{equation}
which corresponds to the quantum extension of the Curzon-Ahlborn efficiency~\cite{Curzon75,Abah12} where the lower classical thermal energy $k_{B}T_{a}$ has been replaced by the ground state energy of a quantum oscillator of frequency $-\Delta_{f}$. As  discussed in the analysis of other proposed quantum heat engines~\cite{RoBFnagel14,Alicki14} this limit, as well as the Carnot limit which in our case is  $[1+\hbar\Delta_{f}/(2k_{B}T)]$, may be surpassed by using a squeezed phonon reservoir or entangled photon and phonon reservoirs.

So far we have considered a quantum heat engine operating on the lower polaritonic branch of the system. The situation is different if we consider the upper polaritonic branch instead: in that case the total work of the optomechanical heat engine is negative, a consequence of the fact that  $\bar{N}_{i}<\bar{N}_{f}$. It follows that if both branches are significantly populated, the effect of the two different cycles counterbalance each other and the total work is reduced. In order to avoid this situation, we had to choose an initial condition that suppresses the thermal population on branch $A$. This was implicitly achieved by starting from a detuning $\Delta_i$ for which the lower polariton branch is strongly phononlike---and hence the upper branch is strongly photonlike--and an initial thermal equilibrium state where the phonon bath is much warmer than the photon bath. At the start the state of the engine is therefore very asymmetrical between photons and phonons, with $\langle\hat{B}^{\dagger}\hat{B} \rangle \gg \langle\hat{A}^{\dagger}\hat{A} \rangle$. However, at stage $(2)$ the situation is reversed and complete thermalization of the system would lead to $\langle\hat{A}^{\dagger}\hat{A} \rangle \gg \langle\hat{B}^{\dagger}\hat{B} \rangle$. Preventing this exchange of populations requires $\gamma\ll1/\tau_{2} < \kappa$, so that the system thermalizes with the cavity reservoir but that process is too fast to have a significant effect on the thermal phonon population. Combined with our previous considerations, the hierarchy of time scales  required for the operation of the proposed heat engine is
\begin{equation}
1/\tau_{4}  <  \gamma \ll 1/\tau_{2} < \kappa < 1/\tau_{1,3} \ll G \ll \omega_{m}.\label{eq:condition}
\end{equation}
As an example, a mechanical resonator of frequency $\omega_m=2\times10^8$~Hz and quality factor $Q = 10^5$, coupled to an optical cavity of linewidth $\kappa=10^6$~Hz and a steady-state occupation of $|\alpha|^2 = 10^{10}$ via a optomechanical coupling $g=10^2$~Hz would fulfill the conditions (\ref{eq:condition}) necessary to realize the proposed Otto cycle \cite{highqoptical}.

To conclude, we return to the assumption that the temperature of the optical bath is essentially $T=0$, so that the phonon bath is by default the ``hot'' reservoir. This is an excellent approximation in the optical regime, but needs not be so in general. Specifically, in the microwave regime the $2.7$~K cosmic blackbody background results in significant photon occupation numbers around $10^2$~GHz frequencies. By the same token, it is also possible to realize quantum mechanical oscillators that operate essentially at $T=0$, for instance in ultracold atomic gases \cite{DSK,Brennecke08}. This suggests that it should be possible to exchange the roles of photons and phonons in our optomechanical heat engine, provided that the mechanical oscillator is cold enough~\cite{Eisert12}. A key condition in that case is that the temperature of the atomic system must be low enough that thermal motion does not wash out the coherent momentum recoil $2\hbar k$ of the atoms due to their interaction with photons of wave vector $k$.  As an example, for a condensate of lithium atoms this condition results in a temperature of the atomic sample not to exceed a pK for $2\pi\times300$ GHz microwave photons. While challenging, this does not seem to be completely impossible.  If realized, a quantum heat engine operating on the upper polariton branch of Fig.~1 would therefore be able to extract heat energy from the cosmic microwave background~\footnote{A promising  alternative approach involving two electromagnetic fields has recently been suggested by Ph. Treutlein (private communication).}.

Future work will carry out detailed dynamical calculations to evaluate the role of imperfections due to the coupling to the thermal reservoirs during all steps of the cycle, with particular emphasis on nonadiabatic transitions between the polariton branches and to nonideal aspects of the control loop required to maintain the mean intracavity power as the detuning is varied. Deep in the quantum regime care is also needed to assess effects such as the conversion of the mean field into polaritons. In that regime the work produced by the engine is extremely small and its detection nontrivial, with measurement backaction expected in general to significantly impact the Otto cycle. 

We thank E. M. Wright and H. Seok for helpful discussions. This work was supported by the National Basic Research Program of China under Grant No. 2011CB921604, the NSFC under Grants No.~11204084 and 11234003, the Specialized Research Fund for the Doctoral Program of Higher Education No.~20120076120003, by SCST under Grant No.~12ZR1443400, the DARPA QuASAR and ORCHID programs through grants from AFOSR and ARO, the U.S. Army Research Office, and the US NSF.


\begin{thebibliography}{10}
\bibitem{note} Note that this is different from the usual thermodynamic definition of adiabaticity, which relates to a transformation without heat exchange.
\end{thebibliography}
\end{document}